\documentstyle[aps,twocolumn,prl,epsf]{revtex}
\begin{document}
\twocolumn[\hsize\textwidth\columnwidth\hsize\csname @twocolumnfalse\endcsname
\title{ Superconducting Instability in the Periodic Anderson Model}

\author{A.\ N.\  Tahvildar-Zadeh,  M.\ H.\ Hettler, M.\ Jarrell\\ }
\address{
         Department of Physics,
University of Cincinnati, Cincinnati, OH 45221-0011\\
 	}

\date{\today}
\maketitle
\widetext
\begin{abstract}
\noindent Employing a quantum Monte Carlo simulation
we find a pairing instability  in the normal state of the
infinite dimensional periodic Anderson model.
Superconductivity arises from a normal
state in which the screening is protracted and which is clearly
not a Fermi liquid.  The phase diagram is reentrant
reflecting competition between superconductivity and Fermi liquid
formation. The estimated superconducting order parameter
is even, but has nodes as a function of
frequency. This opens the possibility of a
 temporal node and an effective order parameter composed of
charge pairs and spin excitations.
\end{abstract}


]

\narrowtext

\paragraph*{Introduction}

A number of Heavy Fermion materials display highly unusual
superconductivity (Ott 1994, Grewe and Steglich 1991, Hess, Riseborough and
Smith 1993). It seems unlikely that conventional
superconductivity coexists with the strong local Coulomb correlations
necessary to enhance the electronic mass in these systems.  Indeed,
the specific heat jump at the transition scales with the normal state
specific heat making it clear that pairing is between the heavy electrons.
However, strong Coulomb correlations present no problem for unconventional
superconducting order parameters with either spatial (Sauls 1994) or
temporal (Berezinski 1974, Balatsky  and Abrahams 1992, Abrahams 1995) nodes.
The latter can occur with either odd or
even-frequency order parameters, which are interpreted in terms of
composite condensates (Bonca and Balatsky 1994).  Such an interpretation is
supported by
two sets of data: (i) power laws observed in physical properties
below the superconducting transition (Ott 1994, Sauls 1994), which
contrast with the activated behavior of the conventional (nodeless)
$s$-wave order of, e.g.,  lead; (ii) the complex superconducting
phase diagrams of UPt$_3$ and U$_{1-x}$Th$_x$Be$_{13}$ (and possibly
UBe$_{13}$ itself, for which the penetration depth displays evidence
of a secondary transition) (Ott 1994, Sauls 1994).  Finally, at least
in UBe$_{13}$ (Ott 1987)  and CeCu$_2$Si$_2$ (Steglich 1996), the
superconductivity arises in a normal state which is clearly
{\it not} a Fermi liquid.  For  $T \agt T_c$ in each of these materials, the
linear specific heat ($C/T$) rises with decreasing $T$ while the resistivity
decreases and has a high residual value at $T_c$
(typically 80-100 $\mu \Omega$-cm in the best samples of UBe$_{13}$).

We have recently analyzed some static (Tahvildar-Zadeh, Jarrell and  Freericks
1997a) and dynamic  (Tahvildar-Zadeh, Jarrell and  Freericks 1997b)
properties of two theoretical paradigms of heavy fermion materials:
the single-impurity Anderson model (SIAM) and the periodic Anderson
model (PAM).    We found for metallic
systems with an $f$-band filling $n_f\approx 1$ and $d$-band filling
$n_d \alt 0.8$, the Kondo scale for the PAM, $T_0$, is strongly
suppressed compared to $T_0^{SIAM}$, the Kondo scale for a
SIAM with the same model parameters.  Consequently, the temperature dependence
of the Kondo peak in the $f$ spectral function is protracted in the PAM,
i.e.\ it is much weaker than in the SIAM, consistent with experiments
on single-crystalline Kondo lattice materials (Andrews 1996).
  However the high
temperature ($T > T_0^{SIAM}$) properties of the two models are
similar, so that $T_0^{SIAM}$ is still the relevant scale for
the onset of screening in the PAM whereas $T_0$, which bears no
obvious relation to $T_0^{SIAM}$, is the scale
for the onset of coherence where the
moment screening is almost complete and the Fermi liquid
begins to form.

We interpreted our results as indicating the emergence of a single
heavy band described by an effective Hubbard model for the local
screening clouds introduced by Nozieres (Nozieres 1985).
Since the effective model is close to half filling  and its hopping 
constant is strongly suppressed by the overlap
of the screened and unscreened states of the f-electron
moments, its Kondo scale, $T_0$, becomes much less than
$T_0^{SIAM}$. In this paper, we show that the regime of
protracted screening is also associated with the occurrence
of a pairing instability in the PAM.\\

{\it Formalism}
The PAM Hamiltonian on a $D$-dimensional hypercubic lattice is,
\begin{eqnarray}
H &=& \frac{-t^*}{2\sqrt{D}}\sum_{\langle ij\rangle \sigma}
\left ( d^\dagger_{i\sigma}d_{j\sigma}+{\rm H.c.}\right )\nonumber \\
&+&
\sum_{i\sigma}\left(
\epsilon_{d}d^\dagger_{i\sigma}d_{i\sigma}+
\epsilon_{f}f^\dagger_{i\sigma}f_{i\sigma}
\right)
+V\sum_{i\sigma}\left(d^\dagger_{i\sigma}
f_{i\sigma}+{\rm H.c.}\right)\nonumber \\
&+&\sum_{i} U(n_{fi\uparrow}-1/2)(n_{fi\downarrow}-1/2)\;\;\label{Ham}.
\end{eqnarray}
where $d(f)^{(\dagger)}_{i\sigma}$ destroys (creates) a $d(f)$
electron with spin $\sigma$ on site $i$. The hopping is restricted
to the nearest neighbors and scaled as $t^*/2\sqrt{D}$.
$U$ is the screened on-site Coulomb repulsion for the localized $f$
states and $V$ is the hybridization between $d$  and  $f$ states.
We choose $t^*$ as the unit of energy ($t^*\approx $
a few electron-volts, the typical band-width of conduction
electrons in metals).

We solve this model Hamiltonian with the dynamical mean field (DMF) method
of Metzner and Vollhardt  (Metzner and Vollhardt 1989)  who observed that the
irreducible
self-energy and vertex-functions become purely local as the coordination
number of the lattice increases. As a consequence, the solution of an
interacting lattice model in $D=\infty$ may be mapped onto the solution of
a local  correlated impurity coupled to a self-consistently determined
host (Pruschke, Jarrell and Freericks 1995; Georges, Kotliar, Krauth and
Rozenberg 1996).   We employ the quantum Monte Carlo (QMC) algorithm
of Hirsch and Fye (Hirsch and Fye 1989)  to solve the remaining impurity
problem and
calculate the  imaginary time, local, one and two particle  Green's functions.
The maximum entropy method (Jarrell and Gubernatis 1996) can then be used to
find the $f$
and $d$ density of states  and  the  self-energy.

\begin{figure}[t]
\epsfxsize=3.375in
\epsffile{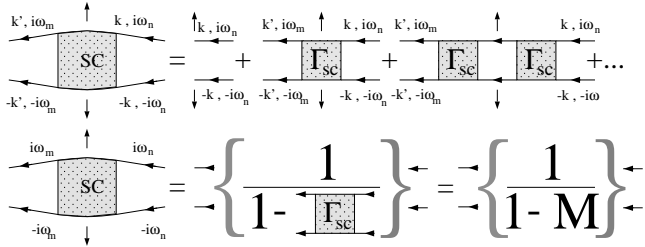}
\caption[a]{Owen and Scalapino's pairing matrix formalism (Owen and Scalapino
1971).
The diagrams for the pair-field susceptibility (top) after integration
over the internal momenta may be re-written (bottom) in terms of the pairing
matrix ${\bf{M}}= \sqrt{\chi^0} \Gamma_{sc} \sqrt{\chi^0}$ (Eq. 2).
Here $\Gamma_{sc}$ is the irreducible two-particle self energy matrix
and $\chi^0$ the bare pairing susceptibility.
The square root is indicated by displaying only ``half'' of the electron
Green's functions in the appropriate places.
Second order phase transitions are signaled
when the largest eigenvalue of ${\bf{M}}$
reaches unity. The corresponding eigenvector $\Phi$ yields information
about the superconducting order parameter.}
\label{SC_diagrams}
\end{figure}

	We search for pairing between the f-electrons by calculating the
f-electron pair-field susceptibility in the normal state.  As shown in
Fig.~\ref{SC_diagrams},
this may be re-expressed in terms of the pairing matrix ${\bf M}$
following Owen and Scalapino  (Owen and Scalapino 1971). ${\bf M}$
is formed from the irreducible superconducting (particle-particle, 
opposite-spin), local vertex function
${\bf \Gamma}_{sc}$, and the bare pair susceptibility
$\chi^0(i \omega_n,{\bf q})= \sum_{i \omega_m,{\bf k}} G(i \omega_m,{\bf k})
G(i \omega_n-i \omega_m,{\bf q-k})$, with $G(i \omega_m,{\bf k})$ the fully
dressed lattice propagators;
\begin{equation}
{\bf{M}}= \sqrt{\chi^0(i \omega_n,{\bf q})}\Gamma^{n,m}_{sc}
\sqrt{\chi^0(i \omega_m,{\bf q})}\,.
\end{equation}
The total pair susceptibility in terms of ${\bf M}$ is given by
\begin{equation}
P^{sc}(T)=\sum_{n,m}
\sqrt{\chi^0(i \omega_n,{\bf q})}\left(1-{\bf{M}}\right)^{-1}
\sqrt{\chi^0(i \omega_m,{\bf q})}\,.
\end{equation}
For the usual second order normal--superconducting transition
$T_c$ is obtained when the largest eigenvalue $\lambda$ of ${\bf M}$
reaches one, so that the pair susceptibility diverges.
The corresponding eigenvector $\Phi(i\omega_n)$ yields information
about the superconducting order parameter.  The pair-field susceptibility
in this most singular channel
can be projected out, using an appropriate frequency form factor
$f_n=\Phi(i\omega_n)/\sqrt{\chi^0({i\omega_n,\bf q})}$,
\begin{equation}
P_\lambda(T)=\sum_{n,m} f_n
\left[\sqrt{\chi^0(\bf q)}\left(1-{\bf{M}}\right)^{-1} \sqrt{\chi^0(\bf
q)}\right]_{n,m}
f_m\,.
\end{equation}

\paragraph*{Results and Interpretation}

\begin{figure}[t]
\epsfxsize=3.375in
\epsffile{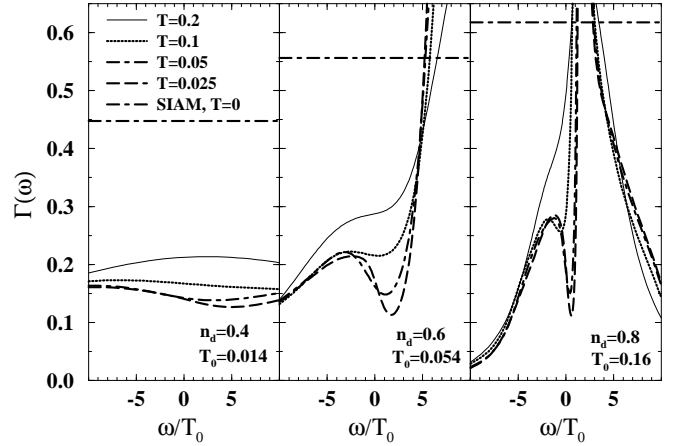}
\caption[a]{Near-Fermi-energy ($\mu=0$) structure of hybridization function
for the asymmetric PAM. The model parameters are $U=1.5$, $V=0.6$ and
$n_f\approx 1$. The dip at the Fermi-energy denotes a decrease in the
number of states available for screening, leading to a suppressed
``coherence'' Kondo scale $T_0$. The zero temperature hybridization for
the single impurity model is shown for comparison.}
\label{Hybrid}
\end{figure}

We introduce an effective hybridization strength $\Gamma(\omega)$,
\begin{equation}
\Gamma(\omega) =
{\rm{Im}}\left(\Sigma(\omega)+{1\over G_f(\omega)}\right),
\end{equation}
where $G_f(\omega)$ is the local $f$ Green's function and
$\Sigma(\omega)$ is the local f-electron self-energy.
$\Gamma(\omega)$ is a measure of the hybridization between the effective
impurity in the DMF problem and its medium (for example, in the SIAM
$\Gamma(\omega)= \pi V^2 N_d(\omega)$, where $N_d(\omega)$
is the d-band density of states).  Fig.~\ref{Hybrid} shows the
effective hybridization for the PAM near the Fermi surface $\mu=0$.
In this figure the model parameters are chosen to be $U=1.5$,
$V=0.6$, $n_f\approx 1$ and three different $d$ band fillings
$n_d=0.4$ ($T_0=0.014$), $n_d=0.6$ ($T_0=0.054$) and $n_d=0.8$
($T_0=0.16$). A relatively small value of $U/V^2$ was chosen
for presentation purposes; however,
the features shown are also present for larger values of $U/V^2$.
(For the parameter set of Fig.~\ref{Hybrid} we do not observe
any superconducting instability, although the hybridization functions look
qualitatively very similar to the case where the instability exists).

The feature we want to emphasize on
is the dip in $\Gamma(\omega)$ at the Fermi energy which develops as the
temperatures is lowered. Since only the electronic states within about
$T_0$ of the Fermi surface participate in the screening (Nozieres 1985)
this indicates a reduced number of available screening states at the Fermi
energy, and since the dip becomes narrower as $n_d\to 1$, the effect is
more dramatic for small $n_d$.   Thus, as the temperature is lowered and
$n_d<0.9$, the Kondo scale of the DMF effective impurity problem is
self-consistently suppressed and hence the ``coherence'' Kondo scale $T_0$ is
also suppressed. Concomitant with this suppression, the Kondo length scale
$l_K\sim 1/T_0$ of the Kondo spin and charge correlations will 
increase dramatically.  We have seen previously that these correlations
can induce ferromagnetism in the PAM (Tahvildar-Zadeh {\it et. al.} 1997a)  and
superconductivity in the
two-channel Kondo model (Jarrell, Pang and Cox 1997).

\begin{figure}[t]
\epsfxsize=3.375in
\epsffile{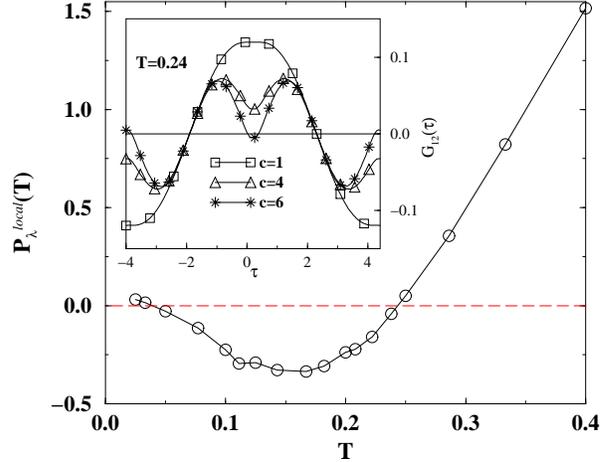}
\caption[a]{
The local pair susceptibility of the dominant eigenvalue versus temperature
for the PAM with $U=2$, $V=0.5$, $n_d=0.6$ and $n_f\approx 1$.  The
susceptibility $P_\lambda^{local}(T)$ is positive in the temperature
regimes $T>T^*_u$ and $T < T^*_l$.  It goes smoothly
through zero at $T^*_{u/l}$ and would be negative in between, indicating
an instability in the corresponding pairing channel. The eigenvalue
$\lambda$ diverges and changes sign at $T^*_{u/l}$ (not shown).
The corresponding eigenvector has both positive and negative parts.
The inset shows the corresponding superconducting order parameter
$G_{f12}(\tau)$ calculated at $T^*_u\approx 0.24$ as described in the
text.  Since $\phi_n$ has arbitrary normalization, we multiply it by a
parameter $c$.  As $c$ increases, the order parameter
develops a suppression at $\tau=0$. This indicates that the two elements
of the pair are not on the same site at the same time, minimizing
the Hubbard energy of the condensate.}
\label{susfig}
\end{figure}

As discussed in the previous section,
a second order superconducting transition is indicated
when the largest eigenvalue of the pairing matrix exceeds unity.
However, in our calculation, the largest positive eigenvalue does not 
approach unity for any value of $U$ or temperature studied. Instead,
for relatively large values of $U\agt 2t^*$ we see a 
different type of instability:
At a high temperature ($T > t^*$) we find generally that all
eigenvalues are either small in magnitude or large and negative.
As the temperature is lowered, the positive eigenvalues grow slowly;
however, the most negative eigenvalue diverges at a temperature 
$T^*_u$.  Upon
lowering the temperature further, the dominant eigenvalue
(that with the largest absolute value) switches sign and becomes
large ($\gg 1$) and positive. It then decreases for some range of
temperatures, but eventually  increases again, diverges  at a 
temperature $T^*_l$, and switches back to large and negative.
The corresponding pair susceptibility $P_\lambda (T)$ goes continuously
through zero at the temperatures $T^*_{u(l)}$ of the divergence of
the dominant eigenvalue  $\lambda$.  $P_\lambda (T)$  is {\it negative} for
$T^*_l < T < T^*_u$, indicating a pairing instability of the
system in this channel. We find that these instabilities are degenerate
throughout the Brillouin zone, i.e., $P_\lambda (T=T^*_{u(l)})$ vanishes  for
every ${\bf q}$ over the whole zone. We demonstrate this by plotting the
corresponding {\it local} pair susceptibility
$P_{\lambda}^{local}(T)$ in Fig.~\ref{susfig} and observe that the
instability prevails even at the local level. Hence this is a
{\it locally} driven transition, consistent with a transition
driven by local dynamical correlations such as those
responsible for Kondo screening. 
We suspect that the degeneracy in the Brillouin zone 
will be lifted in a finite $D$ calculation
by non-local dynamical correlations, such as spin waves, as has been 
suggested previously (Jarrell, Pang and Cox 1997). 

To understand how the vanishing of the pair susceptibility
indicates a phase transition, remember that the inverse pair
susceptibility is proportional to the curvature of the free
energy $f(\Delta)$ as a function of the pair order parameter
$\Delta$, $1/P_\lambda(T) \propto d^2f(\Delta)/d\Delta^2$.
Thus, if $P_\lambda(T)<0$ the normal state becomes
thermodynamically unstable.  The associated transition cannot
be continuous, since this would require $f(\Delta)$ to become
flat for small $\Delta$ (i.e.\  $P_\lambda(T)$ diverges)
so that the order parameter may change continuously.  Thus
the observed transition is {\em{discontinuous}}.
Furthermore, if $P(T^*_u)=0$, then $T^*_u$ is a lower bound
to the transition temperature  since when
$P_\lambda (T-T^*_u=0^-)=0^-$  the curvature of $f(\Delta)$
is divergent and negative, i.e. the free energy displays an
upward cusp. This would compel the order parameter and
the free energy to change discontinuously at $T^*_u$, which
involves an infinite energy at the transition. Hence,
the actual transition occurs at a temperature $T_c>T^*_u$.
Similar arguments can be made to show that  $T^*_l$ is an
upper bound to the discontinuous transition back to the normal
metal.

The dominant eigenvector of the pairing matrix at $T=T^*$ is even in
frequency but has both positive and negative parts. To show the
effect of the nodes on the order parameter in the ordered phase, we
first introduce the Nambu-Gorkov matrix form for the local f-Green's function
\begin{equation} 
{\bf{G}}_f(\omega_n)=\sum_{\bf{k}}\frac{1}
{i\omega_nZ_n{\bf I} - \epsilon_f{\bf \tau}_3-\phi_n{\bf \tau}_1
 - \frac{V^2}{\left[ i\omega_n{\bf{I}}-(\epsilon_k+\epsilon_d){\bf{\tau}}_3
\right]}}
\end{equation}
where ${\bf I}$, ${\bf{\tau}}_1$ and ${\bf{\tau}}_3$ are Pauli matrices.
$Z_n=1-Im(\Sigma(i\omega_n))/\omega_n$ is the wave function renormalization factor
and $\phi_n$ is the renormalized gap function. 
The f-electron order parameter is given by $G_{f12}(\tau)$,
the Fourier transform of the off-diagonal component of ${\bf{G}}_f(i\omega_n)$.
In order to study this order parameter using our knowledge of the normal
state, we assume that $\phi_n$ has the same temporal parity as the dominant eigenvector of the pairing matrix ${\bf M}$ right at $T=T^*$ (in the normal phase). 
But we are not able to predict its actual value by this assumption. 
Hence, as the simplest approximation, we assume that $\phi_n=c f_n$ where
$c$ is an unknown real normalization factor and $f_n$
is the form factor constructed from the dominant eigenvector of the pairing 
matrix at $T=T^*$.
As $c$ increases, $G_{f12}(\tau)$ develops a suppression at $\tau=0$ 
(inset of Fig.~\ref{susfig}).  An order parameter with $G_{f12}(0)\approx 0$
corresponds to a condensate which excludes the simultaneous occupation
of the same site by two electrons, minimizing its Hubbard energy.  For 
the large values of $U\agt 2 t^*$ where an instability is observed, we 
therefore assume that the condensate with $G_{f12}(0)\approx 0$ is
the most stable one. For small $\tau$, 
an order parameter corresponding to an even-frequency $\phi_n$ takes the
form $G_{f12}(\tau) \approx \Delta_0 +\frac12\Delta_2\tau^2$.  If
$\Delta_0=\left<f(\tau=0^+) f(\tau=0)\right>=0$, then
the superconducting condensate is characterized by
$\Delta_2 \propto \left<[[f,H],H]f\right>$.  In the strong-coupling limit,
where the system can be mapped onto the  Kondo lattice model,
the commutators bring  spin  operators into the average (Bonca and Balatsky
1994),
and if $\Delta_0=0$, the order parameter is effectively
a composite of charge pairs and spin-excitations.
Such an order parameter is compelling as it relates
naturally to the complex phase diagram of the superconducting heavy--fermion
materials, in which the competition and sometimes coexistence of
antiferromagnetism and superconductivity is observed. However, it is unclear
whether our estimate corresponds to the actual order parameter of the broken
symmetry state.  A Monte Carlo simulation in the broken symmetry state is
presently underway.\par
\begin{figure}[t]
\epsfxsize=3.375in
\epsffile{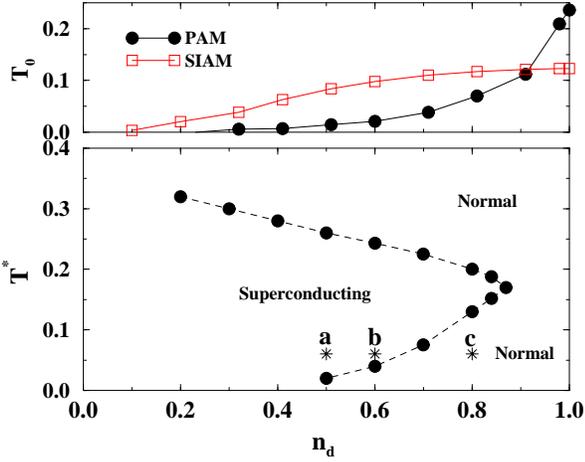}
\caption[a]{Superconducting phase-diagram versus the d-band
filling $n_d$ (bottom) for the PAM with $U=2$, $V=0.5$ and
$n_f\approx 1$. The symbols denote the upper and lower bounds for the
region where the leading eigenvalue of the pairing matrix ${\bf{M}}$
diverges (see text). The dashed line is a guide to the eye.
The upper part shows
the Kondo scales for the PAM and SIAM with the same parameters as for
the lower part.  Note that superconductivity only occurs where the
lattice scale is suppressed relative to the impurity scale, and that
the upper instability temperature $T^*_u>T_0^{PAM}$, indicating that the
superconductivity emerges from a state which is not a Fermi liquid.}
\label{SCphase}
\end{figure}

Fig.~\ref{SCphase} shows the resulting superconducting phase-diagram for
the PAM with parameters $U=2$, $V=0.5$ and $n_f\approx 1$. The symbols
correspond to the temperature where the pairing susceptibility $P_\lambda(T)$
crosses zero and the largest diverging eigenvalue changes sign as discussed
above. Note that the  instabilities cease for $n_d \agt 0.9$,
leading to a phase diagram which suggests  re-entrance into
the normal state. Since $T^*_l$ is only an upper bound to  $T_c^l$,
an actual re-entrance transition need not occur, i.e. $T_c^l$ could be
vanishing. On the other hand,
this would require the upper transition temperature
(which is bounded from below by $T^*_u$)
to drop from a rather large value to zero upon a minute increase of $n_d$.
Although we cannot rule this out numerically,
we believe this scenario to be unlikely. The upper part
of Fig.~\ref{SCphase} shows the Kondo scale for the PAM and SIAM with the
same model parameters as stated above. We see that the  superconducting
instability occurs only when $T_0 < T_0^{SIAM}$, i.e., in the
``protracted screening'' regime. This again suggests that the
pairing  mechanism is related to underscreened moments  and Kondo physics.
Note that the upper transition temperature is large, as is often the case for
a mean field theory. We expect the non-local dynamical
correlations of a finite dimensional system to significantly reduce our
estimate for $T^*_u$. Also, as noted previously (Tahvildar-Zadeh, Jarrell and
Freericks, 1997) the protracted screening regime and the associated 
superconductivity vanish when the orbital f--degeneracy diverges. Thus,
these phenomena are associated with finite orbital degeneracy.

The shape of the lower phase boundary of the transition back to the
normal metal is consistent with the $n_d$ dependence of the
``coherent'' Kondo scale $T_0$.  This indicates that with the
complete screening of the local moments below $T_0$ the system either
looses the mechanism driving superconductivity, or that the system can
gain more free energy by forming a Fermi liquid.  The shape of the
upper phase boundary is more difficult to understand.  $T^*_u$
increases upon lowering $n_d$, in contrast to the simultaneous
decrease of the impurity Kondo scale $T_0^{SIAM}$.  Hence, the onset
of local moment screening alone (described roughly by $T_0^{SIAM}$)
does not lead to superconducting correlations. As observed
by Nozieres (Nozieres 1985), there are not enough states near the
Fermi surface to completely screen all of the f-moments, especially
as the d-band filling decreases.  It is compelling to relate $T^*_u$
to the temperature where the conduction band states available for
screening are exhausted.  Then, for $T<T^*_u$, correlated lattice
effects, such as superconductivity or the magnetic polarons described
by Nozieres, are required to quench the relatively large entropy
associated with the unscreened moments.  Better
understanding of these correlations will be essential in a more
quantitative explanation of the observed superconducting phase.
\par
\begin{figure}[t]
\epsfxsize=3.375in
\epsffile{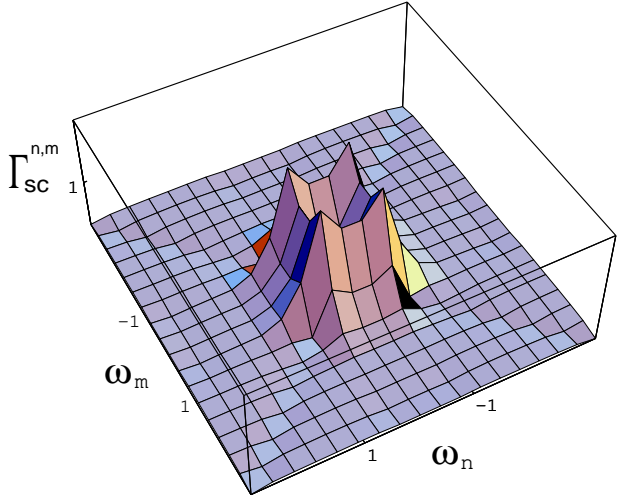}
\caption[a]{The superconducting vertex function corresponding to the point
{\bf b} of Fig.~\ref{SCphase}. The $z$ axis is in units 
of the function value at the central minimum. Note that the vertex function 
is large only at low frequencies, with a width 
$\omega_n \approx 2 \pi T^*_u$.
 }
\label{Vertex}
\end{figure}

Some understanding of these correlations is provided by examining
the Matsubara frequency structure of the vertex function
$\Gamma_{sc}(i\omega_n,i\omega_m)$ which is {\em{minus}} the effective
interaction (e.g.\ to lowest order in perturbation theory 
$\Gamma_{sc}(i\omega_n,i\omega_m)\approx -U$).
Fig.~\ref{Vertex} shows $\Gamma_{sc}$ corresponding to the point {\bf b}
in the phase diagram of Fig.~\ref{SCphase} ({\bf b} is just inside the superconducting region). 
$\Gamma_{sc}(i\omega_n,i\omega_m)$ is large and positive, indicating
an attractive interaction, only at small frequencies, 
and it displays a central minimum.
The width of  the peak at each $n_d$ is roughly $2 \pi T^*_u$, 
which we have argued above to be related to Nozieres' ``exhaustion'' energy scale. Thus, the energy scale inferred from the susceptibility data is 
also present in the vertex function.
This again hints to  Kondo/Nozieres screening as the mechanism 
driving the superconductivity.\\

\begin{figure}[t]
\epsfxsize=3.375in
\epsffile{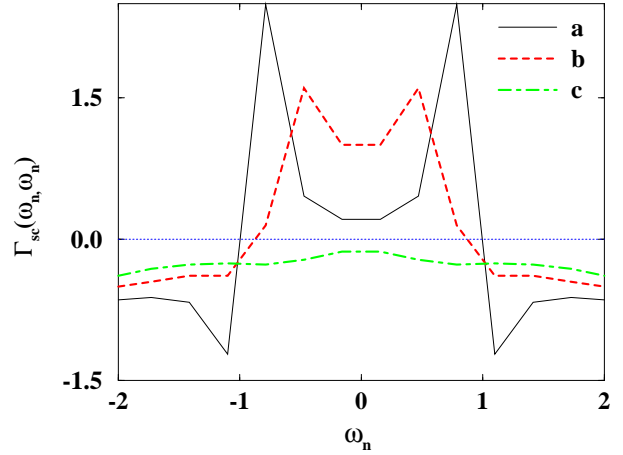}
\caption[a]{The superconducting vertex function in $\omega_n=\omega_m$
direction at three different points of the phase diagram 
(cf. Fig.~\ref{SCphase}). The 
vertex function is purely negative for point {\bf c} outside the 
superconducting region whereas it has positive values at low frequencies  for points {\bf a} and  {\bf b} inside the superconducting region, illustrating
the attractive pair interaction in this region. Note that the narrow peak 
width and the dip around zero frequency 
implies strong retardation of the interaction. 
 }
\label{DiagVertex}
\end{figure}

Fig.~\ref{DiagVertex} shows the irreducible superconducting vertex function
in the $\omega_n=\omega_m$ direction ($\Gamma_{sc}(i\omega_n,i\omega_n)$)
at three different points of the phase diagram (cf. Fig.~\ref{SCphase}). 
We generally find that the vertex function is purely negative for points  outside the superconducting region such as {\bf c}, whereas it has positive values at low frequencies for points inside the superconducting region
such as  {\bf a} and  {\bf b}. This means that the effective interaction
between particles of opposite spin is {\em attractive} for the points
inside the  superconducting phase, but it is {\em repulsive} in
the normal phase.  Furthermore, the narrow width and the dip in the vertex 
function around zero frequency signifies that the {\em attractive} interaction is highly retarded.  The central dip almost
becomes a node at points well within the superconducting phase (such
as {\bf a}) indicating that the static component of the attractive interaction
is not significant.
Therefore, the system would not gain energy by pairing two electrons
on the same site at the same time.  This is consistent with a node 
in the order parameter $G_{f12}(\tau)$ at $\tau = 0$ as discussed above. 

\paragraph*{Conclusions}
In conclusion, we have found a pairing instability in the
infinite-dimensional periodic Anderson model which we interpret
as a superconducting instability.  This instability has
several unusual features including: (i) A reentrant phase diagram
reflecting competition between superconductivity and Fermi liquid
formation; (ii) For most of the phase diagram,
the upper transition temperature $T^*_u\gg T_{0}$, indicating that
superconductivity arises from a normal state which is clearly not
a Fermi liquid; (iii) Although even in frequency, the dominant eigenvector
of the pairing matrix has nodes, which opens the possibility of a
composite order parameter with a temporal node; (iv)
The superconducting instability occurs only in the protracted screening
regime, where the effective hybridization diminishes and the Kondo
length increases with decreasing temperature; (v) Within the superconducting
regime, the vertex function displays a narrow feature with central node 
at low frequencies, consistent with a highly retarded interaction with
a weak static component.  These features, together with
the fact that the instability occurs simultaneously over the whole
zone (i.e., is locally driven), suggests that the pairing
mechanism has its origins in the spin and charge correlations associated
with Kondo/Nozieres screening of the f-electron spins.

	We would like to acknowledge stimulating conversations with
A.\ Arko,
A.V.\ Balatsky,
A.\ Chattopadhyay,
D.L.\ Cox,
J.K.\ Freericks
H.R.\ Krishnamurthy,
T.\ Pruschke,
P.G.J.\ van Dongen
and
F.C.\ Zhang, and in particular D.W.\  Hess.
This work was supported by NSF grants DMR-9704021 and DMR-9357199.
Computer support was provided by the Ohio Supercomputer Center.

\end{document}